\documentclass[a4paper,11pt]{llncs}
\usepackage[a4paper,margin=1in]{geometry}

\usepackage{ifdraft}
\ifdraft{
  \input{vc}
  \usepackage[rgb,x11names]{xcolor}
  \usepackage{prelim2e}
  
}

\usepackage{amsmath}
\usepackage[noend]{algorithmic}
\usepackage{algorithm}
\usepackage{booktabs}
\usepackage{tikz}
\usetikzlibrary{calc,matrix,positioning}

\newcommand{\wc}{\ensuremath{\mathtt{\phi}}}
\newcommand{\sent}{\ensuremath{\mathtt{\$}}}
\newcommand{\bwt}[1]{\ensuremath{{#1}^\mathtt{BWT}}}

\newcommand{\etal}{\textit{et al.,}}
\newcommand{\cf}{\textit{c.f.}}
\newcommand{\eg}{\textit{e.g.,}}
\newcommand{\ie}{\textit{i.e.,}}

\DeclareMathOperator*{\argmin}{arg\,min}
\DeclareMathOperator*{\argmax}{arg\,max}

\begin{document}

\pagestyle{headings}  
\mainmatter              
\title{Succincter Text Indexing with Wildcards}
\titlerunning{Wildcard Support for Succinct Text Indexes}
%
\author{Chris Thachuk}
\authorrunning{C. Thachuk} 
\institute{
Department of Computer Science, 
University of British Columbia, 
Vancouver, Canada
\\\email{cthachuk@cs.ubc.ca}}

\maketitle              

\begin{abstract}
We study the problem of indexing text with wildcard positions,
motivated by the challenge of aligning sequencing data to large
genomes that contain millions of single nucleotide polymorphisms
(SNPs)---positions known to differ between individuals.  SNPs
modeled as wildcards can lead to more informed and biologically
relevant alignments.  We improve the space complexity of previous
approaches by giving a succinct index requiring $(2 + o(1))n \log
\sigma + O(n) + O(d \log n) + O(k \log k)$ bits for a text of length
$n$ over an alphabet of size $\sigma$ containing $d$ groups of $k$
wildcards.  A key to the space reduction is a result we give showing
how any compressed suffix array can be supplemented with auxiliary data
structures occupying $O(n) + O(d \log \frac{n}{d})$ bits to also
support efficient dictionary matching queries.  The query algorithm
for our wildcard index is faster than previous approaches using
reasonable working space.  More importantly our new algorithm greatly
reduces the query working space to $O(d m + m \log n)$ bits.  We note
that compared to previous results this reduces the working space by
two orders of magnitude when aligning short read data to the Human
genome.
\end{abstract}
\section{Introduction}
The study of strings, their properties, and associated algorithms has
played a key role in advancing our understanding of problems in areas
such as compression, text mining, information retrieval, and pattern
matching, amongst numerous others.  A most basic and widely studied
question in stringolgy asks: given a string $T$ (the text) does it
contain a string $P$ (the pattern) as a substring?  It is well known
that this problem can be solved in time proportional to the lengths of
both strings~\cite{KnuMorPra1977}.  However, it is often the case that
we wish to repeat this question for many different pattern strings and
a fixed text $T$ of length $n$ over an alphabet of size $\sigma$.  The
idea is to create a full-text index for $T$ so that repeated queries
can be answered in time proportional to the length of $P$ alone.  It
was first shown by Weiner~\cite{weiner1973} in 1973 that the suffix
tree data structure could be built in linear time for exactly this
purpose.  The ensuing years have seen the versatility of the suffix
tree as it has been demonstrated to solve numerous other related
problems.

While suffix trees use $O(n)$ words of space in theory, this does not
translate to a space efficient data structure in practice.  For this
reason, Manber and Myers~\cite{ManMye1990} proposed the suffix array
data structure (see Figure~\ref{fig:pat-int}).  Though a great practical
improvement over suffix trees, the $\Omega(n \log n)$ bit space
requirement is often prohibitive for larger texts.  Building in part
on the pioneering work of Jacobson~\cite{Jacobson1989} into succinct
data structures, two seminal papers helped usher in the study of
so-called succinct full-text indexes.  Grossi and
Vitter~\cite{GroVit2000} proposed a compressed suffix array that
occupies $O(n \log \sigma)$ bits; the same space required to represent
the original string $T$.  Soon after, Ferragina and
Manzini~\cite{FerMan2000} proposed the FM-index, a type of compressed
suffix array that can be inferred from the Burrows-Wheeler transform
of the text and some auxiliary structures, leading to a space
occupancy proportional to $n H_k(T)$ bits, where $H_k(T)$ denotes the
$k^\text{th}$ order empirical entropy of $T$.  These and subsequent
results have made it possible to efficiently answer the substring
question on texts as large, or larger, than the Human genome.

We are interested in designing a succinct index to answer a
generalized version of the substring question where the text $T$
contains $k$ wildcard positions that can match any character of a
pattern.  Our motivation arises in the
context of aligning short read data, produced by second generation
sequencing technology.  Typically short reads are aligned against a
so-called reference genome; however, the quantity of positions known
to differ between individuals due to single nucleotide polymorphisms
(SNPs) numbers in the millions~\cite{FrazerEtAl2007}. Modeling SNPs as
wildcards would yield more informed, and by extension, more accurate
alignment of short reads.

Cole, Gottlieb \& Lewenstein~\cite{ColGotLew2004} were one of the
first to study the problem of indexing text sequences containing
wildcards and proposed an index using $O(n \log^k n)$ words of space
capable of answering queries in $O(m + \log^k n \log \log n + occ)$
time.  This result was later improved by Lam \etal~\cite{LamEtAl2007}
resulting in space usage of only $O(n)$ words and a query time no
longer exponential in $k$.  A key idea in their work was to build a
type of dictionary of the text segments of $T=T_1 \wc^{k_1} T_2
\wc^{k_2} \ldots \wc^{k_d} T_{d+1}$ where each text segment $T_i$
contains no wildcards and $\wc^{k_i}$ denotes the $i^\text{th}$
\textit{wildcard group} of size $k_i \geq 1$, for $1 \leq i \leq d
\leq k$.  The query time includes the term $\gamma = \sum_{i,j}
\mathsf{prefix}(P[i..|P|],T_j)$ where $\mathsf{prefix}(P[i..|P|],T_j)=1$ if $T_j$
is a prefix of $P[i..|P|]$ and $0$ otherwise.  The authors also give a more
detailed bound on $\gamma$ based on prefix complexity.

Despite this improvement, $O(n)$ words of space is prohibitive for
texts as large as the Human genome.  Support for dictionary matching
of text segments was also crucial in the approach of Tam
\etal~\cite{TamEtAl2009} who proposed the first, and to our knowledge
only, succinct index.  They designed a dictionary structure using $(2
+ o(1)) n \log \sigma$ bits, based on a compressed suffix array, which
therefore occupies most of the space required by their overall index.
Very recently, Belazzougui~\cite{Belazzougui2010} proposed a
succincter dictionary based on the Aho-Corasick automaton having
optimal query time.  The compressed space occupancy was further
improved by a slight modification given by Hon
\etal~\cite{HonEtAl2010}.  While these results are impressive, the
wildcard matching problem benefits from an index that can report the
text segments contained in $P$ (dictionary problem), as well as the
text segments which are prefixed by $P$ and also fully contain $P$.
To draw a distinction, we will refer to this latter type as a
\textit{full-text dictionary}.  In our first main contribution we show
how a full-text dictionary can be built on top of any compressed
suffix array using an additional $O(n) + O(d \log \frac{n}{d})$ bits
of space, and in turn how it can be used to provide a succincter index
for texts containing wildcards.  We note that our dictionary does not
require any modification of the original string $T$.

In our view, the main challenge that must be overcome for successful
wildcard matching is a reduction of the query working space.  The
fastest solution of Tam \etal~\cite{TamEtAl2009}, matches our query
time, if modified to use the same orthogonal range query structure we use, but
requires a query working space of $O(n \log d + m \log n)$ bits.
Acknowledging that the first term is impractical for large texts, they
give a slower solution that reduces the working space to be
proportional to the index itself.  This makes the solution feasible,
but constraining considering the fact that $p$ parallel queries
necessarily increases the working space by a factor of $p$.  In our
second main contribution we give an algorithm that reduces the query
working complexity significantly to $O(d m + m \log n)$ bits.  For our
motivating problem, alignment of short reads (32-64 bases) to the
Human genome (3 billion bases with 1-2 million SNPs), this reduces the
working space by two orders of magnitude from gigabytes to tens of
megabytes.  Our result for indexing text with wildcards is summarized
and compared with existing results in
Table~\ref{tab:wildcard-summary}.
\vspace{-.2in}

\begin{table}\label{tab:wildcard-summary}
\scriptsize
\begin{center}
\begin{tabular}{|p{4cm}|p{6cm}|p{3.25cm}c|}
  \hline
  \textbf{Index Space}
  & \textbf{Query Time}
  & \textbf{Query Working Space}
  &
  \\
  \hline

  $O(n \log^{k} n)$ \hfill words
  & $O(m + \log^k n \log \log n + occ)$
  & - 
  & \cite{ColGotLew2004}
  \\
  \hline
  
  $O(n)$ \hfill words
  & $O(m\log n + \gamma + occ)$
  & $O(n)$ \hfill words
  & \cite{LamEtAl2007}
  \\
  \hline
    
  \begin{minipage}[t]{\linewidth}
   $\left(3+o(1)\right) n \log \sigma$\hfill\\
    $\text{ }\hfill+ O(d\log n) \text{ bits}$
  \end{minipage}
  & 

  \begin{minipage}[t]{\linewidth}
    $O\left(
    \begin{array}{l}
      m\left(\log \sigma + \min\left(m,\hat{d}\right)\log d\right) \\
      + occ_1\log n + occ_2 \log d + \gamma
    \end{array}
    \right)$
  \end{minipage}

  & $O(n \log d + m \log n)$ \hfill bits
  & \cite{TamEtAl2009} 
  \\
  \hline

  \begin{minipage}[t]{\linewidth}
    $\left(3+o(1)\right)n\log \sigma$\hfill\\
    $\text{ }\hfill+ O(d\log n) \text{ bits}$
  \end{minipage}
  & 
  \begin{minipage}[t]{\linewidth}
    $O\left( 
    \begin{array}{l}
      m\left(\log \sigma + \min\left(m,\hat{d}\right)\log d\right)\\
      + occ_1\log n + occ_2 \log d + \gamma \log_\sigma d
    \end{array}
    \right)$
  \end{minipage}

  & $O(n \log \sigma + m \log n)$ \hfill bits
  & \cite{TamEtAl2009}
  \\
  \hline

  \begin{minipage}[t]{\linewidth}
    $(2 + o(1))n \log \sigma + O(n)$\hfill\\
    $\text{ }\hfill+ O(d\log n) + O(k\log k)\text{ bits}$
  \end{minipage}
  & 
  \begin{minipage}[t]{\linewidth}
    $O\left( 
    \begin{array}{l}
      m\left(\log \sigma 
      + \min\left(m,\hat{d}\right)\tfrac{\log k}{\log \log k}\right)\\
      +occ_1 \log n + occ_2 \frac{\log k}{\log \log k} + \gamma
    \end{array}
    \right)$
  \end{minipage}
  & $O(d m + m \log n)$ \hfill bits
  & $\dagger$   
  \\

  \hline
  
\end{tabular}\\
\end{center}
\caption{A comparison of text indexes supporting wildcard characters.
$k$, $d$, $\hat{d}$ is the \# of wildcards, wildcard groups, and
distinct wildcard group lengths, respectively; 
$occ_1, occ_2, occ$ is the \# of Type 1, Type 2, and overall
occurrences, respectively; $\gamma = \sum_{i,j}\mathsf{prefix}(P[i..|P|],T_j)$, $\dagger$ = our result}
\end{table}

\section{Preliminaries}\label{sec:prelim}
Let $T[1,n]$ be a string over a finite alphabet
$\Sigma$ of size $\sigma$.  We denote its $j^\text{th}$ character by
$T[j]$ and a substring from the $i^\text{th}$ to the $j^\text{th}$
position by $T[i..j]$.  
We assume that an end-of-text sentinel character
$\sent \notin \Sigma$ has been appended to $T$ ($T[n]=\sent$) and
\sent{} is lexicographically smaller than any character in $\Sigma$.
For any substring $X$ we use $|X|$ to denote
its length and $\overline{X}$ to denote its reverse sequence.  The
suffix array $\mathsf{SA}$ of $T$ is a permutation of the integers
$[1,n]$ giving the increasing lexicographical order of the suffixes of
$T$.  
Conceptually $\mathsf{SA}$ can be thought of as a matrix of all
suffixes of $T$ that have been sorted lexicographically and where
$\mathsf{SA}[i]=j$ means that the $i^\text{th}$ lexicographically
smallest suffix of $T$ begins position $j$. 

A string $X$ has a suffix array (SA) range $[a,b]$ with respect
to $\mathsf{SA}$ if $a-1$ ($n-b$) suffixes of $T$ are
lexicographically smaller (larger) than $X$.  If $a > b$ the range is
said to be empty and $X$ does not exist as a substring of $T$;
otherwise, $X$ occurs as a prefix of the $b-a+1$ suffixes of $T$
denoted by its range.  The SA range for $X$ can be found in
a compressed suffix array by backwards search using the LF-mapping
which relates $\mathsf{SA}$ to $\bwt{T}$, the Burrows-Wheeler
transform of $T$. $\bwt{T}$ is also a string of length $n$ where
$\bwt{T}[i]=T[\mathsf{SA}[i]-1]$, if $\mathsf{SA}[i] \neq 1$, and
$\bwt{T}[i]=\sent$ otherwise.  See Figure~\ref{fig:pat-int} for an
example.  For details of backwards search, the LF-mapping, existing
implementations, and related topics we refer the reader to the
excellent review by Navarro and M{\"a}kinen~\cite{NavMak2007}.  In
this work, we assume the availability of a compressed suffix array
meeting the following space and time requirements, of which there are
many (\cf{}~\cite{NavMak2007}).

\begin{lemma}\label{lem:csa-bounds}
A compressed suffix array $\mathsf{SA}$ for $T$ can be represented in
$(1 + o(1))n \log \sigma$ bits of space, such that the suffix
array range of every suffix of a string $X$ can be computed in $O(|X|
\log \sigma)$ time, and each match of $X$ in $T$ can be reported in an
additional $O(\log n)$ time.
\end{lemma}

\noindent In our dictionary construction, we also make use of the
following well known data structures.

\begin{lemma}[Raman \etal{}~\cite{RamanEtAl2002}]\label{lem:bit-vec}
  A bit vector $\mathsf{B}$ of length $n$ containing $d$ 1 bits can be
  represented in $d \log \frac{n}{d} + O(d + n \frac{\log \log n}{\log
    n})$ bits to support the operations $\mathtt{rank}_1(\mathsf{B},
  i)$ giving the number of 1 bits appearing in $\mathsf{B}[1..i]$ and
  $\mathtt{select}_1(\mathsf{B}, i)$ giving the position of the
  $i^\text{th}$ 1 in $\mathsf{B}$ in $O(1)$ time.
\end{lemma}

\begin{lemma}[Grossi \& Vitter~\cite{GroVit2000}]\label{lem:comp-int-array}
An array $\mathsf{L}$ of $d$ integers where $\sum_{i=1}^{d}
{\mathsf{L}[i]} = n$ can be represented in $d(\lceil \lg(n/d) \rceil +
2 + o(1))$ bits to support $O(1)$ time access to any element.
\end{lemma}

\begin{lemma}[Munro \& Raman~\cite{MunRam2002}]\label{lem:bp-rep}
  A sequence $\mathsf{BP}$ of $d$ balanced parentheses can be
  represented in $(2+o(1))d$ bits of space to support the following
  operations in $O(1)$ time: $\mathtt{rank}_((\mathsf{BP}, i)$,
  $\mathtt{select}_((\mathsf{BP}, i)$, and similarly for right
  parentheses, as well as:
  \begin{itemize}
    \item $\mathtt{findclose}(\mathsf{BP}, l)$: index of matching right
      parenthesis for left parenthesis at position $l$
    \item $\mathtt{enclose}(\mathsf{BP},i)$: indexes $(l,r)$ of
      closest matching pair to enclose
      $(i,\mathtt{findclose}(\mathsf{BP}, i))$ if such a pair exists
      and returns an undefined interval in $\mathsf{BP}$ otherwise
  \end{itemize}
\end{lemma}

The matching statistics for a string $X$ with respect to $\mathsf{SA}$
is an array $ms$ of tuples such that $ms[i] = (q, [a,b])$ states that
the longest prefix of $X[i..|X|]$ that matches anywhere in $T$ has
length $q$ and suffix array range $[a,b]$.  Very recently Ohlebusch
\etal{}~\cite{OhlebuschEtAl2010} showed matching statistics can be
efficiently computed with backward search if $\mathsf{SA}$ is
\textit{enhanced}
with auxiliary data structures using $O(n)$ bits to represent
so-called longest common prefix intervals
(\cf{}~\cite{OhlebuschEtAl2010}).
We leverage this result in the design of our succinct full-text
dictionary and its search algorithm.

\begin{lemma}[Ohlebusch \etal{}~\cite{OhlebuschEtAl2010}]\label{lem:ms-alg}
  The matching statistics of a pattern $X$ with respect to text $T$
  over an alphabet of size $\sigma$ can be computed in $O(|X|\log
  \sigma)$ time given a compressed enhanced suffix array of $T$.
\end{lemma}

\noindent Finally, our wildcard matching algorithm makes use of an orthogonal
range query data structure.

\begin{lemma}[Bose \etal{}~\cite{BoseEtAl2009}]\label{lem:range-query}
  A set $N$ of points from universe $M = [1..k] \times
  [1..k]$, where $k = |N|$, can be represented in $(1 + o(1))k
  \log k$ bits to support orthogonal range reporting in $O(occ
  \frac{\log k}{\log \log k})$ time, where $occ$ is the size of the
  output.
\end{lemma}

\section{A succinct full-text dictionary}\label{sec:dictionary}

In the dictionary problem we are required to index a set of $d$ text
segments\footnote{To remain consistent with the section that follows we
  refer to dictionary entries (patterns) as text segments.}
$\mathcal{D}=\{T_1,T_2,\ldots,T_d\}$ so that we can efficiently match
in any input string $P$ all occurrences of text segments belonging to
$\mathcal{D}$.  We present a succinct \textit{full-text dictionary}
index that is also capable of efficiently identifying all text
segments that contain $P$ as a prefix, or more generally as a
substring.  We demonstrate the use of this additional functionality in
our solution for wildcard matching.

\subsection{A compressed suffix array representation of text segments}

Let $T = \wc T_1 \wc T_2 \wc T_3 \wc \ldots \wc T_d \sent$ be the
concatenation of all $d$ text segments, each prefixed by the character
\wc, followed by the traditional end-of-text sentinel \sent, having
total length $n$.  Note that $n$ is necessarily larger than the total
number of character in the dictionary.  We define \wc{} to be
lexicographically smaller than any
$c \in \Sigma$ and \sent{} to be lexicographically smaller than \wc.
We first build $\mathsf{SA}$, the compressed suffix array for $T$.
Consider any text segment $T_j \in \mathcal{D}$.  There will be a
contiguous range $[c,d]$ of suffixes in $\mathsf{SA}$ that are
prefixed by the string $T_j$.  Lemma~\ref{lem:patt-in-text} summarizes
how we can use the SA range of $T_j$ and its length to determine if it
is prefix of a given text $P$ (and vice versa).

\begin{lemma}\label{lem:patt-in-text}
Let $\mathsf{SA}$ be the compressed suffix array for $T$ and let
$[a,b]$ and $[c,d]$ be the non-empty suffix array ranges in
$\mathsf{SA}$ for a string $P$ and a text segment $T_j$ respectively.
Then $T_j$ is a prefix of $P$ if and only if $c \leq a \leq b \leq d$
and $|P| \geq |T_j|$.  Similarly, $P$ is a prefix of $T_j$ if and only
if $a \leq c \leq d \leq b$.
\end{lemma}

\subsection{Storing text segment lengths}

For Lemma~\ref{lem:patt-in-text} to apply, we must know both the SA
range of a given text segment and also its length.  By
Lemma~\ref{lem:comp-int-array} we can store the lengths of all $d$
text segments in a compressed integer array $\mathsf{L}$ using
$d(\lceil \log (n/d)\rceil + 2 + o(1))$ bits ensuring constant time
access.  We store the lengths in $\mathsf{L}$ relative to the
lexicographical order of text segments.

\subsection{The text segment interval tree}

\newcounter{ycount}
\setcounter{ycount}{0}

\begin{figure}[t]
\begin{tikzpicture}
  [x=0.4cm, y=0.4cm,
  pat int/.style={outer sep=2pt, inner sep=0, fill=white, 
    font=\footnotesize, anchor=east,},
  ]
  
  \tikzset{node style ge/.style={rectangle}}
  \matrix (A) [matrix of math nodes, 
    anchor=base east,
    nodes = {
      node style ge,
    },
    row sep=-2mm,
    column sep=0mm,
    column 1/.style={anchor=base west},
    column 3/.style={anchor=base east},
    column 4/.style={anchor=base east},
    column 5/.style={anchor=base east},
    column 6/.style={anchor=base east},
  ] { 
    & \bwt{T} & \mathsf{SA} & \mathsf{B} & i \\
    \mathtt{\sent                                             } & \mathtt{c    } & 21  & 0 &  1 \\
    \mathtt{\wc a \wc aa \wc cacc \wc ac \sent                } & \mathtt{a    } &  8  & 0 &  2 \\
    \mathtt{\wc aa \wc aca \wc a \wc aa \wc cacc \wc ac \sent } & \mathtt{\sent} &  1  & 0 &  3 \\
    \mathtt{\wc aa \wc cacc \wc ac \sent                      } & \mathtt{a    } & 10  & 0 &  4 \\
    \mathtt{\wc ac \sent                                      } & \mathtt{c    } & 18  & 0 &  5 \\
    \mathtt{\wc aca \wc a \wc aa \wc cacc \wc ac \sent        } & \mathtt{a    } &  4  & 0 &  6 \\
    \mathtt{\wc cacc \wc ac \sent                             } & \mathtt{a    } & 13  & 0 &  7 \\
    \mathtt{a \wc a \wc aa \wc cacc \wc ac \sent              } & \mathtt{c    } &  7  & 1 &  8 \\
    \mathtt{a \wc aa \wc cacc \wc ac \sent                    } & \mathtt{\wc  } &  9  & 0 &  9 \\
    \mathtt{a \wc aca \wc a \wc aa \wc cacc \wc ac \sent      } & \mathtt{a    } &  3  & 0 & 10 \\
    \mathtt{a \wc cacc \wc ac \sent                           } & \mathtt{a    } & 12  & 0 & 11 \\
    \mathtt{aa \wc aca \wc a \wc aa \wc cacc \wc ac \sent     } & \mathtt{\wc  } &  2  & 1 & 12 \\
    \mathtt{aa \wc cacc \wc ac \sent                          } & \mathtt{\wc  } & 11  & 0 & 13 \\
    \mathtt{ac \sent                                          } & \mathtt{\wc  } & 19  & 1 & 14 \\
    \mathtt{aca \wc a \wc aa \wc cacc \wc ac \sent            } & \mathtt{\wc  } &  5  & 1 & 15 \\
    \mathtt{acc \wc ac \sent                                  } & \mathtt{c    } & 15  & 1 & 16 \\
    \mathtt{c \sent                                           } & \mathtt{a    } & 20  & 1 & 17 \\
    \mathtt{c \wc ac \sent                                    } & \mathtt{c    } & 17  & 0 & 18 \\
    \mathtt{ca \wc a \wc aa \wc cacc \wc ac \sent             } & \mathtt{a    } &  6  & 0 & 19 \\
    \mathtt{cacc \wc ac \sent                                 } & \mathtt{\wc  } & 14  & 1 & 20 \\
    \mathtt{cc \wc ac \sent                                   } & \mathtt{a    } & 16  & 1 & 21 \\
  };

  \foreach \from/\to/\id/\label/\offsetya/\offsetyb/\offsetx in
           {9/17/1/a/0/-.1/-6.1,    
            13/14/2/aa/.2/.1/-3.5,  
            13/14/3/aa/0/.25/-.5,   
            15/17/4/ac/-.1/.1/-3.5, 
            16/16/5/aca/0/.1/-.5,   
            21/21/6/cacc/0/0/-.5    
           } {    
    \coordinate (A-\from-\to-from) at ($(A-\from-1.north west)+(0,-.4)+(0,\offsetya)$);
    \coordinate (A-\from-\to-to) at ($(A-\to-1.south west)+(0,.2)+(0,\offsetyb)$);
    \coordinate (A-\from-\to-center) at ($(A-\from-\to-from) !.5!
    (A-\from-\to-to) + (\offsetx, 0)$);

    \draw[blue] (A-\from-\to-from)
    -| (A-\from-\to-center) node[pat int] {$(\id,\mathtt{\label})$} 
    |- (A-\from-\to-to);
  };

  \matrix [matrix of math nodes,
    anchor=west,
  ] at ($(A.east)+(0.5,0)$)
  { 
    \mathsf{BP} & ( & ( & ( & ) & ) & ( & ( & ) & ) & ) & (  & ) \\
    \mathsf{L}  & 1 & 2 & 2 & 2 & 3 & 4 \\
    \mathsf{R}  & 0 & 0 & 2 & 2 & 3 & 5 & 5 & 6 \\
               i & 1 & 2 & 3 & 4 & 5 & 6 & 7 & 8 & 9 & 10 & 11 & 12 \\
  };
 
\end{tikzpicture}
\caption{A succinct full-text dictionary for the set of text segments
  $\{\mathtt{aa, aca, a, aa, cacc, ac}\}$.  Shown are the sorted
  suffixes of the string $T=\mathtt{\wc aa \wc aca \wc a \wc aa \wc
    cacc \wc ac\sent}$ representing the text segments.  Text segment
  intervals are demarcated on the left and labeled by their
  lexicographical order (lex id) and the text segment they represent.}
\label{fig:pat-int}
\end{figure}

The SA range of one text segment $T_i$ will enclose the SA range of
another $T_j$ if $T_i$ is a prefix of $T_j$.  For instance, in the
example of Figure~\ref{fig:pat-int} the text segment $\mathtt{aca}$
has SA range $[15,15]$ and is enclosed by the SA range of the text
segment $\mathtt{ac}$ ($[14,16]$) and by the text segment $\mathtt{a}$
($[8,16]$).  In general, it is also possible that many text segments
begin at the same position, provided that they are different
occurrences of the same string (\eg{} $\mathtt{aa}$).  This is by
design since each text segment is followed by a character not found in
$\Sigma$ (either \wc{} or \sent{}).  However, our construction
requires us to distinguish between different occurrences of the same
text segment string and we therefore introduce the concept of
\textit{text segment intervals}.  When $t > 1$ text segments in the
dictionary share a common SA range we say that the text segment
interval of occurrence $a$ encloses the text segment interval of
occurrence $b$, $1 \leq a \neq b \leq t$, if the suffix of $T$
beginning with occurrence $a$ is lexicographically smaller than the
suffix beginning with occurrence $b$.  In this way we are able to
define a total order on all $d$ text segment intervals based on their
relative lexicographical order in $\mathsf{SA}$.  We assign
\textit{lex ids}, a unique identifier for each text segment, based on
this lexicographical order.  Consider again the example in
Figure~\ref{fig:pat-int}.  The text segment $\mathtt{aa}$ occurs as a
prefix of $T[2..n]$ and $T[11..n]$.  Since the suffix $T[2..n]$ is
lexicographically smaller than $T[10..n]$, we say that the occurrence
prefixing $T[2..n]$ encloses the other.  Consequently, the text
segment prefixing $T[2..n]$ ($T[11..n]$) is assigned lex id $2$ ($3$).
We will refer to text segments or text segment intervals
interchangeably.

In general the text segment intervals form a set of nested,
non-crossing intervals (an interval tree) and can be represented by a
sequence $\mathsf{BP}$ of $d$ balanced parentheses; one pair for each
text segment (see Figure~\ref{fig:pat-int}).  Conceptually, if we can
identify the text segment interval having the largest lex id that is a
prefix of $P$, referred to as the \textit{smallest enclosing text
  segment interval} of $P$, then we can immediately conclude that $P$
is also prefixed by all intervals which enclose it.

\begin{lemma}\label{lem:count-report-pat}
Given the index pair $(l,r)$ in $\mathsf{BP}$ corresponding to the
smallest enclosing text segment interval for a string $P$ the $occ$
number of text segments that are prefixes of $P$ can be counted in
$O(1)$ time and reported in an additional $O(occ)$ time.
\end{lemma}

\subsection{Finding the smallest enclosing text segment interval}

We now describe how the smallest enclosing text segment interval can
be determined given any non-empty SA range $[a,b]$ in $\mathsf{SA}$
for $P$.  We wish to determine the pair $(l,r)$ of indexes for the
left and right parentheses in $\mathsf{BP}$ corresponding to this
interval (or an undefined index range if $P$ is not prefixed by any
text segment).  Unfortunately, we cannot directly infer where text
segment intervals begin and end based on $\bwt{T}$ alone.  Therefore,
we make use of a bit vector $\mathsf{B}$ of length $n$ and set
$\mathsf{B}[k]=1$ if and only if one or more text segment intervals
begin at position $k$, or end at position $k-1$.  For the range
$[a,b]$, end cases occur when $\mathsf{B}[k]=0$, $a < k \leq n$ (all
text segment intervals end before position $a$) or when
$\mathsf{B}[k]=0$, $1 \leq k \leq a$ (all text segment intervals begin
after position $a$).  Suppose otherwise and let $c=\argmax_{1 \leq j
  \leq a}\{\mathsf{B}[j]=1\}$ and $d=\argmin_{a < j \leq
  n}\{\mathsf{B}[j]=1\}$.  Note that position $c$ marks the largest
position (up to $a$) when one or more text segment intervals begin or
end (at $c-1$).  Our algorithm considers two main cases: either $B[c]$
marks the beginning of one or more intervals, or it only marks the end
of intervals.

\begin{lemma}\label{lem:B-array}
  Given two positions $c$ and $d$ of $\mathsf{B}$, where $c < d$,
  $\mathsf{B}[c]=\mathsf{B}[d]=1$ and $\mathsf{B}[k]=0$, $c < k < d$,
  then $\mathsf{B}[c]$ marks the beginning of $t$ text segment intervals if
  and only if $\bwt{T}[c..d-1]$ contains $t$ occurrences of the
  character \wc.
\end{lemma}

Using Lemma~\ref{lem:B-array} we are able to distinguish between the
two main cases.  If $\mathsf{B}[c]$ marks the beginning of one or more
text segment intervals, then $T_j$ --- the text segment interval with
the largest lex id beginning at position $c$ --- is the smallest
enclosing text segment interval, provided $|T_j| \leq |P|$ (by
condition of Lemma~\ref{lem:patt-in-text}).  If $|T_j| \leq |P|$, we
can determine the largest lex id beginning at position $c$ by simply
counting the occurrences of the character \wc{} prior to position $d$
in $\bwt{T}$.  Conveniently and by construction, this corresponds to
the rank of the left parenthesis denoting $T_j$ in $\mathsf{BP}$.  It
is worth noting that when $|T_j| > |P|$ special care is required to
find the smallest enclosing text segment interval in worst case
constant time.  Details are given in the proof of
Lemma~\ref{lem:enc-text-seg}, but the idea is to find the enclosing
interval (if any) of the text segment interval having the
\textit{smallest} lex id beginning at position $c$.

On the other hand, if $\mathsf{B}[c]$ only marks the end of one or
more text segment intervals, we can instead identify the right index
for $T_{j'}$ --- the last text segment interval (smallest lex id) to
end at position $c-1$.  The smallest enclosing text segment interval,
if any, is therefore the one enclosing $T_{j'}$.  Unfortunately, in
this case we cannot infer how many intervals close prior to position
$c$ directly from $\bwt{T}$.  For this reason, we will employ another
compressed integer array $\mathsf{R}$ to record the count of intervals
that close prior to position $k$, for all $\mathsf{B}[k]=1$.  We
determine the appropriate index for $\mathsf{R}$ by simply counting
the number of $1$'s up to position $c$ in $\mathsf{B}$.  The
corresponding entry in $\mathsf{R}$ gives us the rank of the right
parenthesis for the last interval to close prior to position $c$, from
which we can find the enclosing interval (if any).  The entire
procedure, including end cases, is summarized in
Algorithm~\ref{alg:enc-text-seg} and shown correct in
Lemma~\ref{lem:enc-text-seg}.

\begin{algorithm}[t]
  \caption{Find smallest enclosing text segment interval}
  \begin{algorithmic}[1]
    \REQUIRE {$a$ specifies the beginning of the non-empty suffix array interval for string $P$}
    \ENSURE {$l,r$ where $l$ ($r$) is the index of the left
      (right) parenthesis in $\mathsf{BP}$ corresponding to the
      smallest enclosing text segment interval of $P$ if it exists,
      and an undefined interval otherwise}
    \STATE $c\gets \mathtt{select}_1(\mathsf{B}, \mathtt{rank}_1(\mathsf{B}, a))$
    \STATE $d\gets \mathtt{select}_1(\mathsf{B}, \mathtt{rank}_1(\mathsf{B}, a) + 1)$
    \IF[handle end cases] {$c$ or $d$ is undefined}
    \RETURN an undefined interval
    \ENDIF
    \STATE $lexid\gets \mathtt{rank}_{\wc{}}(\bwt{T}, d-1)$
    \IF[\ensuremath{\mathsf{B}[c]} marks beginning of t.s. interval(s)] {$lexid > \mathtt{rank}_{\wc{}}(\bwt{T}, c)$}
    \IF {$\mathsf{L}[lexid] > |P|$}
    \STATE $lexid\gets \mathtt{rank}_{\wc{}}(\bwt{T}, c-1) + 1$
    \STATE $l\gets \mathtt{select}_{\mathtt{(}}(\mathsf{BP},lexid)$
    \STATE $l,r\gets\mathtt{enclose}(\mathsf{BP},l)$
    \ELSE
    \STATE $l\gets \mathtt{select}_{\mathtt{(}}(\mathsf{BP}, lexid)$
    \STATE $r\gets \mathtt{findclose}(\mathsf{BP}, l)$
    \ENDIF
    \ELSE[\ensuremath{\mathsf{B}[c]} marks end of t.s. interval(s)]
    \STATE $r\gets \mathtt{select}_{\mathtt{)}}(\mathsf{BP}, \mathsf{R}[\mathtt{rank}_1(\mathsf{B},c)])$
    \STATE $l\gets \mathtt{findopen}(\mathsf{BP}, r)$
    \STATE $l, r\gets \mathtt{enclose}(\mathsf{BP}, l)$
    \ENDIF
    \RETURN $l, r$
  \end{algorithmic}
  \label{alg:enc-text-seg}
\end{algorithm}

\begin{lemma} \label{lem:enc-text-seg}
  Let $\mathsf{SA}$ be the compressed suffix array for $T$ and let
  $[a,b]$ be the non-empty suffix array range in $\mathsf{SA}$ for a
  string $P$.  In $O(1)$ time, Algorithm~\ref{alg:enc-text-seg} either
  correctly identifies the indexes in $\mathsf{BP}$ corresponding to
  the smallest enclosing text segment interval of $P$ if one exists,
  or it returns an undefined interval when it does not.
\end{lemma}

\subsection{The overall dictionary and its full-text capabilities}

We have shown how all text segments occurring as a prefix of a string
$P$ having a non-empty SA range in $\mathsf{SA}$ can be reported
efficiently.  By enhancing $\mathsf{SA}$ with lcp-interval information
using $O(n)$ bits, we can find the matching statistics for $P$ in
order to repeat the previous procedure for $1 \leq i \leq |P|$ (see
Lemma~\ref{lem:ms-alg}).  Importantly for our results on wildcard
matching, we note that with a very minor modification, this same
construction works when text segments are separated by more than one
\wc{} character and also when the first text segment is not preceded
by a \wc{} character.  Note that the text segment interval tree can be
built in a similar manner as an lcp-interval tree.  Details are left
for the full version.  We have our first main result.

\begin{theorem}\label{thm:patt-dict}
  Given a set of $d$ text segments over an alphabet of size $\sigma$
  we can construct a succinct full-text dictionary, based on an
  enhanced compressed suffix array, using at most $(1 + o(1))n \log
  \sigma + O(n) + O(d \log \frac{n}{d})$ bits where $n$ is the length
  of $T$, the text representation of the dictionary including \wc{}
  characters, such that the $\gamma$ text segments contained in a
  string $P$ can be counted in $O(|P|\log \sigma)$ time and reported
  in an additional $O(\gamma)$ time.  Furthermore, all text segments
  prefixed by $P$ can be reported in $O(|P|\log \sigma + occ)$ time,
  and all locations in $T$ where $P$ occurs as a substring can be
  reported in $O(|P|\log \sigma + occ \log n)$ time.
\end{theorem}

\section{Matching wildcards in succinct texts}
Let $T$ be a string over an alphabet $\Sigma \cup \{\wc\}$ of size
$\sigma$ where $\wc \notin \Sigma$ and $T[i]=\wc$ if and only if
position $i$ is a wildcard position in $T$.  In particular, we denote
the structure of the input string as $T=T_1 \wc^{k_1} T_2 \wc^{k_2}
\ldots \wc^{k_d} T_{d+1}$ where each text segment $T_i$ contains no
wildcards and $\wc^{k_i}$ denotes the $i^\text{th}$ \textit{wildcard
  group} of size $k_i \geq 1$, for $1 \leq i \leq d$.  Our goal is to
create an index for the purpose of identifying all the locations in
$T$ that exactly match any query pattern $P$, modulo wildcard
positions.  Similar to previous
approaches~\cite{LamEtAl2007,TamEtAl2009}, we classify the match into
one of three cases: $X$ contains no wildcard group (Type 1), $X$
contains exactly one wildcard group (Type 2), and $X$ contains more
than one wildcard group (Type 3).

\subsection{Overall design of the index}
We first build the succinct full-text dictionary of
Section~\ref{sec:dictionary}.  By design, the dictionary reports the
match of a text segment $T_j$ based on its lexicographical order (its
lex id) relative to other text segments; however, in the wildcard
problem we are required to report the match based on $T_j$'s position
in $T$.  Therefore, we store a permutation $\Pi$ mapping the lex ids
of text segments to their relative position order in $T$.  For
instance, if $T_j$ has lex id $k$, then $\Pi[k]=j$.  We find it
convenient to store the following information for each text segment,
in auxiliary arrays, indexed by this relative position order: length,
SA range in $\mathsf{SA}$ (referenced as $\mathsf{RSA}$), beginning
position in $T$, and the size of the preceding wildcard group.  Note
that array $\mathsf{L}$ of the dictionary construction can be adapted
to store lengths in this relative order with the use of $\Pi$.  We
also construct a compressed suffix array $\overline{\mathsf{SA}}$ for
$\overline{T}$, the reverse of $T$, and store the SA range of each
$\overline{T_j}$ with respect to $\overline{\mathsf{SA}}$ (referenced
as $\overline{\mathsf{RSA}}$).  Note that $\overline{\mathsf{SA}}$
does not need to support location reporting.  We use simple arrays to
store SA ranges resulting in $O(d \log n)$ bits combined space usage
to store auxiliary information supporting constant time access.  To
support Type 2 matching we employ a range query data structure
occupying $(1+o(1))k \log k$ bits (see next section).

\begin{lemma}\label{lem:wildcard-space}
  Given a text $T$ of length $n$ containing $d$ groups of $k$
  wildcards the combined space required of the above indexes is $(2 +
  o(1))n \log \sigma + O(n) + O(d \log n) + O(k \log k)$ bits.
\end{lemma}

All three matching types make use of the matching statistics of $P$
with respect to $\mathsf{SA}$.  Types 2 and 3 matching also make use
of the SA ranges of $\overline{P}$ with respect to
$\overline{\mathsf{SA}}$.  Both can be computed in $O(m \log \sigma)$
time (by Lemmas~\ref{lem:csa-bounds} and \ref{lem:ms-alg}) and require $O(m
\log n)$ bits to store.  We incorporate these times and working space
into the results for each type.  Type 1 matching is handled
by the application of Lemma~\ref{lem:csa-bounds}.

\subsection{Type 2 matching}

A Type 2 match occurs when the alignment of $P$ to $T$ contains
exactly (a portion of) one wildcard group.  Specifically, we seek a
pair of neighbouring text segments $T_j$ and $T_{j+1}$, separated by a
wildcard group of size $k_j$, where $P[i..|P|]$ aligns to the first
$|P|-i+1$ characters of $T_{j+1}$ --- referred to as the
\textit{suffix match} (of $P$) --- and $P[1..i-1-k_j]$ aligns to the
last $i-1-k_j$ characters of $T_j$ --- referred to as the
\textit{prefix match}.  Let $\alpha_j$ ($\omega_j$) be the the first
(last) \wc{} character of the $j^\text{th}$ wildcard group in $T$.
End cases occur when the match begins or ends in
$T[\alpha_j'..\omega_j']$, where $\alpha_j'$ ($\omega_j'$) is the
position of $\alpha_j$ ($\omega_j$) in $T$.  For now, suppose this is
not the case.  For a fixed suffix $P[i..|P|]$ and wildcard group
length $k_j$ our strategy will be to (i) find all potential suffix
matches, (ii) record the lex id of the candidate text segments, (iii)
find all potential prefix matches, and (iv) determine which candidate
prefix matches are compatible with a lex id recorded in step (ii).

\begin{center}
\begin{tikzpicture}
  \node (la) at (0,0) {$\mathtt{\cdots}$};
  \node (lb) at (3,0) {$\wc$};
  \node (lc) at (4,0) {$\wc$};
  \node (ld) at (7,0) {$\mathtt{\cdots}$};
  \draw[-|] (la) -- node[fill=white] {$T_{j}$} (lb);
  \path (lb) -- node[fill=white] {$\mathtt{\cdots}$} (lc);
  \draw[|-] (lc) -- node[fill=white] {$T_{j+1}$} (ld);

  \foreach \id/\label in {lb/$\alpha_{j}$, lc/$\omega_j$} {
    \draw[<-] (\id.center)+(0,-.25) -- +(0,-.75) node[anchor=base,fill=white] {\label};
  }
\end{tikzpicture}
\end{center}

\begin{lemma}\label{lem:text-in-patt}
  Given a non-empty SA range $[a,b]$ in $\mathsf{SA}$ for a string
  $X$, the lex ids (based on their lexicographical order) of text
  segments in $T$ that contain $X$ as a prefix will form a contiguous
  (possibly empty) range $[id_1,id_2]$ that can be reported in $O(1)$
  time.
\end{lemma}

By Lemma~\ref{lem:text-in-patt}, we can identify the range
$[id_1,id_2]$ of lex ids corresponding to text segments that
$P[i..|P|]$ is a prefix of in constant time using its stored SA range
with respect to $\mathsf{SA}$, completing steps (i)-(ii).  Determining
a range $[id_3,id_4]$ of lex ids corresponding to text segments that
$P[1..i-k_j-1]$ is a suffix of is equivalent to determining all
$\overline{T_{t}}$ that contain $\overline{P[1..i-k_j-1]}$ as a
prefix.  Again, using a stored SA range with respect to
$\overline{\mathsf{SA}}$ this can be determined in constant time,
completing step (iii).  Now consider that the lex id with respect to
$\mathsf{SA}$ of a text segment $T_{j+1}$ is relative to the rank of
$\omega_j$ in $\bwt{T}$, the character which precedes it.  Similarly,
the relative rank of $\alpha_j$ in $\overline{\bwt{T}}$ determines the
lex id of $\overline{T_j}$, but in this case relative to
$\overline{T}$.  We make use of a permutation $\mathsf{H}$ to relate
these lex ids ($\alpha$ and $\omega$ values).  Specifically, we set
$\mathsf{H}[\alpha_j]=\omega_j$, for $1 \leq j \leq k$.  Therefore, we
need to determine the entries in $\mathsf{H}[id_3..id_4]$ that have a
value in the range $[id_1,id_2]$.  This is an orthogonal range query
and by Lemma~\ref{lem:range-query}, $\mathsf{H}$ can be represented in
$(1+o(1))k \log k$ bits to report all $occ$ matches in $O(occ
\frac{\log k}{\log \log k})$ time.  Once a lex id $\omega_j$ has been
verified, a match position can be reported in $O(1)$ time as the
location of $T_{j+1}$ with respect to $T$ is known in addition to the
length of the prefix match.  This completes step (iv).

In general, we can repeat the above procedure for every combination of
suffix length and wildcard group length bound by $m$.  However, as
pointed out by Tam \etal~\cite{TamEtAl2009} the number of distinct
wildcard group sizes $\hat{d}$ is often a small constant, particularly
in genomic sequences.  We therefore only consider at most $\hat{d}$
lengths, provided they are not larger than $m$.

Now, consider the case when $P[i..|P|]$ aligns to a prefix of a
wildcard group.  To contain $P[i..|P|]$ as a prefix, the wildcard
group must have a length $l \geq |P|-i+1$.  Let $a$ be the first entry
in $\mathsf{SA}$ denoting a suffix of $T$ prefixed by at least $l-1$
\wc{} characters and let $b$ be the last entry prefixed by any \wc{}
character.  Then, similar to Lemma~\ref{lem:text-in-patt},
$\bwt{T}[a..b]$ will contain a range $[id_1,id_2]$ giving ranks of
\wc{} characters in that interval.  Some sub-sequence of $[id_1,id_2]$
will correspond to $\omega$ wildcards that begin groups having length
$l$ or longer.  Therefore, Type 2 matches can be determined by
reporting entries in $\mathsf{H}[id_3..id_4]$ having a value in
$[id_1,id_2]$, where $[id_3,id_4]$ is defined as before.  The case
when a prefix of $P$ aligns as a suffix of a wildcard group can be
handled similarly.  Note that the SA ranges of the at most $m$
wildcard group lengths we are interested in can be determined in $O(m
\log \sigma)$ time and stored in $O(m \log n)$ bits.

\begin{lemma} \label{lem:type2}
  All Type 2 matches can be reported using $O(m \log n)$ bits of
  working space in $O(m(\log \sigma + \min(m,\hat{d})\frac{\log
    k}{\log \log k}) + occ_2\frac{\log k}{\log \log k})$ time.
\end{lemma}

\subsection{Type 3 matching}

\begin{algorithm}[t]
  \caption{Report Type 3 matches}
  \begin{algorithmic}[1]
    \REQUIRE {a string $P$ of length $m$, its matching statistics
      w.r.t. $\mathsf{SA}$, SA ranges for all suffixes of $\overline{P}$
      w.r.t. $\overline{\mathsf{SA}}$}
    \ENSURE {positions in $T$ matching $P$, modulo wildcard positions}
    \FOR{$i = 1$ \TO $m$} 
    \STATE let $(q,[a,b])$ be the matching statistics for $P[i..m]$
    \STATE use Algorithm 1 to find indexes $(l,r)$ in $\mathsf{BP}$ denoting smallest
    enclosing text segment interval for SA range
    $[a,b]$ 
    \WHILE{$(l,r)$ is a defined interval in $\mathsf{BP}$} 
    \STATE $lex id\gets \mathtt{rank}_((\mathsf{BP}, l)$
    \STATE $j\gets \Pi[lex id]$
    \STATE $[a_p,b_p],\,[a_s,b_s]\gets $ SA range of
    $\overline{P[1..i-1-k_{j-1}]}$ w.r.t $\overline{\mathsf{SA}}$, SA range of $P[i+l_j+k_j..m]$ w.r.t $\mathsf{SA}$
    \STATE $[c_s,d_s],\, [c_p,d_p]\gets \mathsf{RSA}[j-1],\, \overline{\mathsf{RSA}}[j+1]$
    \IF[Case 1: $P$ does not contain $T_{j-1}$]{$i \leq l_{j-1} + k_{j-1}$}
    \IF[Case 1: prefix condition satisfied]{$k_{j-1} \geq i - 1$
      \OR $[a_p,b_p]$ encloses $[c_p,d_p]$}
    \IF[Case 1a: $P$ does not contain $T_{j+1}$]{$m - i + 1 < l_j +
      k_j + l_{j+1} - 1$}
    \IF[Case 1a: suffix condition satisfied]{$m-i \leq l_j + k_j$ \OR
      $[a_s,b_s]$ encloses $[c_s,d_s]$}
    \PRINT match at position $x_j-i+1$
    \ENDIF
    \ELSE[Case 1b: $P$ must contain $T_{j+1}$]
    \STATE set $(i + l_j + k_j)^\text{th}$ bit of $\mathsf{W}[j+1]$ to 1
    \ENDIF
    \ENDIF
    \ELSE[Case 2: $P$ must contain $T_{j-1}$]
    \IF[Case 2: prefix condition is satisfied]{$i^\text{th}$ bit of $\mathsf{W}[j]$ is set to 1}
    \IF[Case 2a: $P$ does not contain $T_{j+1}$]{$m - i + 1 < l_j +
      k_j + l_{j+1} - 1$}
    \IF[Case 2a: suffix condition satisfied]{$m-i \leq l_j + k_j$ \OR
      $[a_s,b_s]$ encloses $[c_s,d_s]$}
    \PRINT match at position $x_j-i+1$
    \ENDIF
    \ELSE[Case 2b: $P$ must contain $T_{j+1}$]
    \STATE set $(i + l_j + k_j)^\text{th}$ bit of $\mathsf{W}[j+1]$ to 1
    \ENDIF
    \ENDIF
    \ENDIF
    \STATE $(l,r)\gets \mathtt{enclose}(\mathsf{BP}, l)$
    \ENDWHILE
    \ENDFOR
  \end{algorithmic}
  \label{alg:type3-match}
  \textbf{Notation}: $x_j,\, l_j, \, k_j$ denotes the position,
  length and wildcard group length (which follows) the text segment $T_j$
\end{algorithm}

Type 3 matches contain at least (portions of) two wildcard groups and
therefore must fully contain at least one text segment.  The general
idea in previous approaches and in this paper is to consider this case
as an extension of the dictionary matching problem: text segments
contained within $P$ are candidate positions, but we must verify if
they can be extended to a full match of $P$.  However, we execute this
idea in an altogether novel manner that greatly reduces the working
space over existing approaches. The complete details of our approach
are given in Algorithm~\ref{alg:type3-match}.  We now highlight the
main idea and give the intuition behind the correctness but note that
a formal proof is given in the appendix.  

First, suppose that text segment $T_j$ matches $P$ starting at
position $i$.  Consider the conditions that must be satisfied to
confirm that this match can be extended to a complete match of $P$ in
$T$.  We must verify that (i) $P[1..i-1]$ can be matched to the text
preceding $T_j$ in $T$ --- referred to as the \textit{prefix
  condition} --- and (ii) $P[i+|T_j|..|P|]$ can be matched to the text
following $T_j$ in $T$ --- referred to as the \textit{suffix
  condition}.  If both conditions are verified, we can report that $P$
matches $T$ at position $x_j-i+1$, where $x_j$ is the start position
of $T_j$ in $T$.

\begin{center}
\begin{tikzpicture}
  \node (la) at (0,0) {$\mathtt{\cdots \wc}$};
  \node (lb) at (3,0) {$\mathtt{\wc \cdots \wc}$};
  \node (ca) at (6,0) {$\mathtt{\wc \cdots \wc}$};
  \node (cb) at (9,0) {$\mathtt{\wc \cdots}$};

  \draw[|-|] (la) -- node[fill=white] {$T_{j-1}$} (lb);
  \draw[|-|] (lb) -- node[fill=white] {$T_{j}$} (ca);
  \draw[|-|] (ca) -- node[fill=white] {$T_{j+1}$} (cb);

  \foreach \id/\label in {la/$x_{j-1}$, lb/$x_j$, ca/$x_{j+1}$} {
    \draw[<-] (\id.east)+(0,-.2) -- +(0,-.75) node[anchor=base,fill=white] {\label};
  }
\end{tikzpicture}
\end{center}

For working space, we make use of an array $\mathsf{W}$ containing
$d+1$ entries (one for each text segment) of $m$ bits, with all
entries set to zero using the constant time initialization
technique~\cite{BriggsTorzcon1993}.  During the course of the
algorithm the $i^\text{th}$ bit of $\mathsf{W}[j]$ is set to 1 if the
prefix condition is true for $P[1..i-1]$ with respect to $T_j$.  There
are exactly $m$ stages of the algorithm $(i=1,\ldots,m$) corresponding
to the suffixes of $P$.  In a given stage $i$ we consider each text
segment $T_j$ found to be a prefix of the $i^\text{th}$ suffix of $P$.
To verify the prefix and suffix conditions for $T_j$ we first consider
(line 9 of Algorithm~\ref{alg:type3-match}): will $P[1..i-1]$ need to
fully contain the previous text segment $T_{j-1}$ in order to match in
$T$?  This breaks our algorithm into the two main cases.  If not (Case
1), we check the prefix condition by checking whether $P[1..i-1]$ is
compatible with the wildcard group to its left and the suffix of
$T_{j-1}$ to which it must align (line 10).  If the prefix condition
is satisfied, we consider (line 11): will $P[i+|T_j|..m]$ need to
fully contain the next text segment $T_{j+1}$ in order to match in
$T$?  If not (Case 1a), we check whether the suffix condition is
satisfied by checking that $P[i+|T_j|..m]$ is compatible with the
wildcard group to its right and the prefix of $T_{j+1}$ to which it
must align (line 12).  If indeed the suffix condition is satisfied, we
output a match (line 13).  If yes (Case 1b), we set the
$(i+l_j+k_j)^\text{th}$ bit of entry $\mathsf{W}[j+1]$ to 1, to
indicate that a prefix condition holds for $P[1..i+l_j+k_j-1]$ with
respect to $T_{j+1}$ (line 15).  The key idea here is that we only
attempt to verify the suffix condition when $T_j$ would be the last
text segment to occur in $P$ (\ie{} Case 1a) and if not (Case 1b), we
record information in $\mathsf{W}$ stating that we currently have a
partial match, but for it to remain viable, $T_{j+1}$ should be a
prefix of $P[i+l_j+k_j..m]$.  Case 2 occurs when $P$ must contain the
previous text segment $T_{j-1}$ to satisfy the prefix condition (lines
16--22).  Since stages of the algorithm proceed with increasing values
of $i$, then the prefix condition would have been previously checked
and, if satisfied, the $i^\text{th}$ bit of $\mathsf{W}[j]$ would be
set to 1.  The remaining questions are answered as before: the suffix
condition is verified if possible, and otherwise successful partial
matches are again recorded in $\mathsf{W}$.

\begin{lemma} \label{lem:type3}
  All Type 3 matches can be reported in $O(m \log \sigma +
  \gamma)$ time using $O(d m + m \log n)$ bits of working space.
\end{lemma}

\noindent Combining the results for the 3 types of matching we arrive
at our second main result.

\begin{theorem} \label{thm:wc-match}
  Given a text $T$ of length $n$ containing $d$ groups of $k$
  wildcards all matches of a pattern $P$ of length $m$ can be reported
  using $O(d m + m \log n)$ bits of working space in $O(m(\log \sigma
  + \min(m,\hat{d})\frac{\log k}{\log \log k}) + occ_1 \log n + occ_2
  \frac{\log k}{\log \log k} + \gamma)$ time with an index occupying
  $(2 + o(1))n \log \sigma + O(n) + O(d \log n) + O(k \log k)$ bits of space.
\end{theorem}

\subsubsection{Acknowledgments.}  The author would like to thank Anne
Condon for helpful discussions, detailed feedback and suggestions on
this manuscript.

\bibliographystyle{splncs03}
\bibliography{cpm2011-cthachuk}

\begin{thebibliography}{10}
\providecommand{\url}[1]{\texttt{#1}}
\providecommand{\urlprefix}{URL }

\bibitem{Belazzougui2010}
Belazzougui, D.: {Succinct dictionary matching with no slowdown}. In: CPM. pp.
  88--100. Springer (2010)

\bibitem{BoseEtAl2009}
Bose, P., He, M., Maheshwari, A., Morin, P.: {Succinct orthogonal range search
  structures on a grid with applications to text indexing}. Algorithms and Data
  Structures pp. 98--109 (2009)

\bibitem{BriggsTorzcon1993}
Briggs, P., Torczon, L.: An efficient representation for sparse sets. ACM Lett.
  Program. Lang. Syst.  2,  59--69 (1993)

\bibitem{ColGotLew2004}
Cole, R., Gottlieb, L.A., Lewenstein, M.: Dictionary matching and indexing with
  errors and don't cares. In: Thirty-sixth annual ACM symposium on Theory of
  computing. pp. 91--100. STOC '04, ACM (2004)

\bibitem{FerMan2000}
Ferragina, P., Manzini, G.: {Opportunistic data structures with applications}.
  In: Foundations of Computer Science, 2000. Proceedings. 41st Annual Symposium
  on. pp. 390--398. IEEE (2002)

\bibitem{FrazerEtAl2007}
Frazer, K., Ballinger, D., Cox, D., Hinds, D., Stuve, L., Gibbs, R., et~al.: {A
  second generation human haplotype map of over 3.1 million SNPs}. Nature
  449(7164),  851--861 (2007)

\bibitem{GroVit2000}
Grossi, R., Vitter, J.: {Compressed suffix arrays and suffix trees with
  applications to text indexing and string matching}. In: {Thirty-second annual
  ACM Symposium on Theory of Computing}. pp. 397--406. ACM (2000)

\bibitem{HonEtAl2010}
Hon, W., Ku, T., Shah, R., Thankachan, S., Vitter, J.: {Faster Compressed
  Dictionary Matching}. In: String Processing and Information Retrieval. pp.
  191--200. Springer (2010)

\bibitem{Jacobson1989}
Jacobson, G.: Succinct static data structures. Ph.D. thesis, Carnegie Mellon
  University (1989)

\bibitem{KnuMorPra1977}
Knuth, D., Morris~Jr, J., Pratt, V.: {Fast pattern matching in strings}. SIAM
  J. on Computing  6,  323 (1977)

\bibitem{LamEtAl2007}
Lam, T.W., Sung, W.K., Tam, S.L., Yiu, S.M.: Space efficient indexes for string
  matching with don't cares. In: Proceedings of the 18th international
  conference on Algorithms and computation. pp. 846--857 (2007)

\bibitem{ManMye1990}
Manber, U., Myers, G.: Suffix arrays: a new method for on-line string searches.
  In: SODA '90: Proceedings of the first annual ACM-SIAM symposium on Discrete
  algorithms. pp. 319--327. Society for Industrial and Applied Mathematics,
  Philadelphia, PA, USA (1990)

\bibitem{MunRam2002}
Munro, J., Raman, V.: {Succinct representation of balanced parentheses and
  static trees}. SIAM Journal on Computing  31(3),  762--776 (2002)

\bibitem{NavMak2007}
Navarro, G., M{\"a}kinen, V.: {Compressed full-text indexes}. ACM Computing
  Surveys (CSUR)  39(1), ~2 (2007)

\bibitem{OhlebuschEtAl2010}
Ohlebusch, E., Gog, S., K\"{u}gel, A.: Computing matching statistics and
  maximal exact matches on compressed full-text indexes. In: {SPIRE}, vol.
  6393, pp. 347--358. Springer (2010)

\bibitem{RamanEtAl2002}
Raman, R., Raman, V., Rao, S.: {Succinct indexable dictionaries with
  applications to encoding k-ary trees and multisets}. In: Proceedings of the
  thirteenth annual ACM-SIAM symposium on Discrete algorithms. pp. 233--242.
  Society for Industrial and Applied Mathematics Philadelphia, PA, USA (2002)

\bibitem{TamEtAl2009}
Tam, A., Wu, E., Lam, T.W., Yiu, S.M.: Succinct text indexing with wildcards.
  In: String Processing and Information Retrieval, vol. 5721, pp. 39--50.
  Springer (2009)

\bibitem{weiner1973}
Weiner, P.: {Linear pattern matching algorithms}. In: 14th Annual Symposium on
  Switching and Automata Theory. pp. 1--11. IEEE (1973)

\end{thebibliography}

\newpage
\appendix
\section{Supporting Proofs}

\subsubsection{Proof of Lemma~\ref{lem:patt-in-text}}

\begin{proof}
  We first consider the case for determining if $T_j$ is a prefix of
  $P$.   Suppose that $T_j$ is a prefix of $P$.  Then it must be the case
  that $|T_j| \leq |P|$.  By definition $T[\mathsf{SA}[c]..|T|]$
  ($T[\mathsf{SA}[d]..|T|]$) is lexicographically smaller (greater) than
  any other suffix of $T$ prefixed by the string $T_j$; thus, $[c,d]$
  must enclose $[a,b]$ and we have $c \leq a \leq b \leq d$.

  Next consider the case when $c \leq a \leq b \leq d$ and $|T_j| \leq
  |P|$.  Since $[c,d]$ encloses $[a,b]$ they must share a common
  prefix of length $\min(|P|,|T_j|)$.  If $[a,b]=[c,d]$ it could be
  the case that $P$ is a proper prefix of $T_j$; however, since $|P|
  \geq |T_j|$ then $P$ and $T_j$ must share a common prefix of length
  at least $|T_j|$.  Thus, $T_j$ is a prefix of $T$.

  The other case is symmetric, but it is not necessary to compare the
  lengths of $P$ and $T_j$.
  \qed
\end{proof}

\subsubsection{Proof of Lemma~\ref{lem:text-in-patt}}

\begin{proof}
  This follows from the proof of
  Lemma~\ref{lem:patt-in-text} and by the definition of lex ids since
  they correspond to \wc{} characters (which prefix text segment
  occurrences only) in \bwt{T} that must necessarily be contained
  within the SA range for those text segment occurrences.\qed
\end{proof}

\subsubsection{Proof of Lemma~\ref{lem:count-report-pat}}
\begin{proof}
  We let $I_1$ denote the interval in $\mathsf{BP}$ specified by
  $(l,r)$.  If $I_1$ is an undefined interval then $P$ is not prefixed
  by any text segment ($occ=0$) and we are done.  Suppose $I_1$ is
  defined.  This interval is enclosed by another interval $I_2=(p,q)$
  if and only if $p < l$ and $q > r$.  Since text segment intervals
  cannot cross, if $I_2$ opens before $I_1$ ($p < l$) it is either the
  case that $I_2$ closes before $I_1$ opens ($q < l$) or $I_2$ closes
  after $I_1$ closes ($q > r$); it is the latter case we are
  interested in.  We count the number of intervals that begin (opening
  parentheses), up to index $l$, and subtract the number which also
  end (closing parentheses), up to index $l$.  The difference is
  exactly the number of enclosing intervals for $I_1$.  Specifically,
  $occ = \mathtt{rank}_\mathtt{(}(\mathsf{BP},l) -
  \mathtt{rank}_\mathtt{)}(\mathsf{BP},l)$ and can be computed in
  $O(1)$ time.

  Reporting the text segment match for interval $I_1$ consists of
  outputting a tuple containing $(start, end, lex id)$.  The $lex id$ is the
  lexicographical order of the text segment (relative to others) and
  is determined in $O(1)$ time as
  $lex id=\mathtt{rank}_\mathtt{(}(\mathsf{BP},l)$.  Since we report text
  segments that are prefixes of $T$, then $start=1$ and
  $end=start+\mathsf{L}[lex id]-1$ (as lengths of text segments are
  stored in $\mathsf{L}$ according to their lex id).  After reporting
  the match for $I_1$, we can determine the next enclosing interval by
  setting $(l,r)=\mathtt{enclose}(\mathsf{BP},l)$ and repeating the
  above procedure until all $occ$ occurrences have been reported.\qed
\end{proof}

\subsubsection{Proof of Lemma~\ref{lem:B-array}}

\begin{proof}
Suppose $t$ text segment intervals begin at position $c$.  As previously
stated, if two or more text segment intervals begin at the same position
then they are different occurrences of the same text segment string
$\omega$.  By definition of $\mathsf{B}$, no other text segment interval
can begin before position $d$ in $\mathsf{SA}$.  If $\mathsf{B}[d]$
marks the beginning of another text segment interval, it must be
lexicographically larger than $\omega$ and therefore all $t$
occurrences of $\omega$ appear before position $d$.  If
$\mathsf{B}[d]$ instead marks the end of one or more text segment intervals
(at position $d-1$), it must be for the $t$ occurrences of $\omega$
since text segment intervals cannot cross.  In either case, all occurrences
of the text segment $\omega$ must appear in $\mathsf{SA}$ in the range
$[c..d-1]$ (possibly in addition to other suffixes of $T$ prefixed by
the string $\omega$).  Since only text segment instances are prefixed by
the character \wc{} in $T$, then $\bwt{T}[c..d-1]$ must contain
exactly $t$ occurrences of \wc.

Suppose $\bwt{T}[c..d-1]$ contains $t$ occurrences of the character
\wc.  Since each text segment occurrence is prefixed by the character \wc{}
in $T$, then $t$ suffixes of $T$ in the range $[c..d-1]$ of
$\mathsf{SA}$ are prefixed by text segment occurrences.  Each text segment
occurrence corresponds to one text segment interval.  Text segment intervals
only begin in positions $k$ where $\mathsf{B}[k]=1$.  Therefore $t$
text segment intervals begin at position $c$ as no other text segment
intervals can begin before position $d$, by definition of
$\mathsf{B}$.\qed
\end{proof}

\subsubsection{Proof of Lemma~\ref{lem:enc-text-seg} (Algorithm 1
  - Find smallest enclosing text segment interval)}

\begin{proof}
  Algorithm~\ref{alg:enc-text-seg} begins by identifying the last entry in
  $\mathsf{B}$ up to position $a$ and the first entry after position
  $a$ equal to $1$ denoting the opening or closing of text segment
  intervals.  If either of these are undefined, then a text segment
  interval cannot enclose $[a,b]$ and an empty interval is returned
  (lines 3--4).

  If $\bwt{T}[c..d-1]$ contains one or more \wc{} characters then by
  Lemma~\ref{lem:B-array}, $\mathsf{B}[c]$ marks the beginning of some
  number of text segment intervals (lines 6--13).  Since text segment
  interval lex ids are based on their lexicographical order in
  $\mathsf{SA}$, then the lex id of the last text segment interval to
  open at position $c$ is $lexid$, given by the count of \wc{} characters
  up to position $d-1$ in $\bwt{T}$.  Let $T_k$ be this text segment
  interval.  By Lemma~\ref{lem:patt-in-text}, we must also ensure that
  $|P| \geq |T_k|$ by checking the text segment length in $\mathsf{L}$
  (line 7).  If $P$ is shorter than $T_k$ (lines 8--10), then it is
  also shorter than all text segment intervals beginning at position
  $c$ since they represent the same text segment string.  However, it
  is possible that there exists a text segment interval $T_j$ that is a
  longest proper prefix of $T_k$.  Note that $|P| > |T_j|$, since it
  must be lexicographically larger than $T_j$; otherwise
  $\mathsf{B}[c]$ would correspond to this interval instead of $T_k$.
  If $T_j$ exists, it would enclose the first text segment interval that
  begins at position $c$.  We can find the lex id for the first text
  segment interval opening at position $c$ (smallest lex id) similarly to $T_k$, but instead we
  count the occurrences of \wc{} prior to position $c$ and then add
  one.  The lex id will correspond to the rank of the left parenthesis in
  $\mathsf{BP}$ and the index $l$ is easily determined by a
  $\mathtt{select}$ operation.  Note that the $\mathtt{enclose}$
  operation will return an undefined interval if $T_j$ does not exist.  If
  instead $|T_k| \leq |P|$ (lines 12--13), we can simply determine the
  index for the left parenthesis denoting the text segment interval $T_k$.

  Otherwise, $\mathsf{B}[c]$ only marks the end of some text segment
  interval(s) (lines 15--17).  In this case, we use the number of
  occurrences of 1's in $\mathsf{B}$ up to position $c$ as an index
  into the array $\mathsf{R}$ which stores the number of text segment
  intervals that close prior to the position denoted by that entry.
  This allows us to identify $T_k$, the last text segment interval to
  close prior to position $a$ (the one having the smallest lex id).
  If another text segment interval $T_j$ encloses $T_k$, then it must be
  the case that $T_j$ encloses $[a,b]$ and $|T_j| < |P|$, otherwise
  $T_j$ would also close prior to position $a$.

  At this point, the pair $(l,r)$ either correctly identifies the
  smallest enclosing text segment interval for the SA range $[a,b]$,
  or it is an undefined interval if none exists.  Overall, a constant
  number of operations are required and all can be computed in $O(1)$
  time.\qed
\end{proof}

\subsubsection{Proof of Lemma~\ref{lem:wildcard-space} (Space analysis
  of our succinct wildcard index)}

\begin{proof}
  The succinct full-text dictionary requires $(1+o(1))n \log \sigma +
  O(n) + O(d \log \frac{n}{d})$ bits by Theorem~\ref{thm:patt-dict},
  which in turn is based on a combination of Lemmas 1--4, and the
  additional $O(n)$ bits required to \textit{enhance} $\mathsf{SA}$
  with lcp-interval information and to store the LCP array.  The
  wildcard index also requires a suffix array of the reverse of string
  $T$ which occupies $(1 + o(1))n \log \sigma$ bits by
  Lemma~\ref{lem:csa-bounds}.  The most space dominant auxiliary array
  is used to store suffix array ranges in $O(d \log n)$ bits.  The
  range query data structure requires $O(k \log k)$ bits by
  Lemma~\ref{lem:range-query}.  Thus,
  overall we have a space complexity of $(2 + o(1))n\log \sigma + O(n)
  + O(d \log n) + O(k \log k)$ bits.  \qed
\end{proof}

\subsubsection{Proof of Lemma~\ref{lem:type3} (Algorithm 2 - Type 3 Matching)}

\begin{proof}
Recall that the algorithm proceeds in $m$ stages for increasing
$i=1,\ldots,m$ for each suffix of $P$.  It is clear in the algorithm
description that verification of a match of $T_j$ proceeds by first
ensuring the prefix condition can be satisfied (Case 1: if $P$ does not
contain $T_{j-1}$) or ensuring it was previously satisfied  (Case 2:
$P$ must contain $T_{j-1}$), and then verifying the suffix condition
in the cases where $P$ does not contain $T_{j+1}$ (Cases 1a, 2a) (and
reporting a match when verified), or by instead marking $\mathsf{W}$
to signify a partial match, expecting the match to be continued by a
match of $T_{j+1}$ at the time step $i+l_j+k_j$ (Cases 1b, 2b).  The
correctness relies on showing that $\mathsf{W}$ is set correctly to
confirm the satisfaction of the prefix condition for the next text
segment ($T_{j+1}$) for a future time step.
We show correctness by induction on $i$.  Consider the base case
($i=1$).  All candidate text segments $T_j$ fall into Case 1 which
(importantly) does not rely on the correctness of previous steps of
the algorithm.  The prefix condition is trivially true.  Thus, if a
successful match of $P[1..m]$ to $T[x_j..n]$ will not fully contain
$T_{j+1}$ we can simply check if $P[l_j+k_j+1..m]$ is a prefix of $T_{j+1}$ by
Lemma~\ref{lem:patt-in-text}.  If it is, both conditions have been
satisfied and we have a match, otherwise, we record in $\mathsf{W}[j+1]$ that
$T_{j+1}$ must appear as a prefix of $P[l_j+k_j+1..m]$ to form a
successful match.  Now assume we are in step $i$ and the algorithm is
correct up to step $i-1$.  Case 1 is handled as before and does not
rely on the correctness of previous steps, so assume we are in Case 2
($P$ must contain $T_{j-1}$).  Then, if the prefix condition is
satisfied the $i^\text{th}$ bit of $\mathsf{W}[j]$ should be set to
1.  Since this would have been set at some step $t < i$, and we have
assumed the algorithm is correct up to step $i-1$, then it must be the
case that the prefix condition for $T_j$ is satisfied if and only
if $\mathsf{W}[j]$ has bit $i$ set to 1.  Similarly to before, if the
prefix condition is satisfied, we can attempt to verify the suffix
condition using Lemma~\ref{lem:patt-in-text} when $P$ does not contain
$T_{j+1}$ or by recording the partial match in $\mathsf{W}$ as
before.  This completes the correctness proof.

We now consider the additional runtime and work space incurred for
Type 3 matching.  There are $\gamma$ candidate positions overall that
can be reported in $O(m \log \sigma + \gamma)$ time by
Theorem~\ref{thm:patt-dict}.  Each candidate is processed once, in
$O(1)$ time.  The array $\mathsf{W}$ occupies $O(d m)$ bits as working
space.  Thus, the overall time complexity is $O(m \log \sigma +
\gamma)$ and working space is $O(d m + m \log n)$.\qed
\end{proof}

\end{document}